\documentclass{aa}
\usepackage{graphicx}
\usepackage{natbib}
\usepackage{amssymb}
\bibpunct{(}{)}{;}{a}{}{,}
\begin{document}
   \title{Probing the inner wind of AGB stars: Interferometric observations of SiO millimetre line emission from the oxygen-rich stars \object{R Dor} and \object{L$^2$ Pup}}

\titlerunning{Probing the inner wind of AGB stars}


   \author{F. L. Sch\"oier\inst{1}  \and H. Olofsson\inst{1} \and T. Wong\inst{2,3} \and M. Lindqvist\inst{4}
   \and F. Kerschbaum\inst{5}}

   \offprints{F. L. Sch\"oier \\ \email{fredrik@astro.su.se}}

   \institute{Stockholm Observatory, AlbaNova, SE-106 91 Stockholm, Sweden
   \and CSIRO Australia Telescope National Facility, PO Box 76, 
Epping NSW 1710, Australia
\and School of Physics, University of New South Wales, Sydney NSW 2052, Australia
   \and Onsala Space Observatory, SE-439 92 Onsala, Sweden
   \and Institut f\"ur Astronomie, T\"urkenschanzstra{\ss}e 17, 1180 Wien, Austria}    
             
   \date{Received; accepted}

\abstract{High angular resolution Australia Telescope Compact Array (ATCA) observations of SiO `thermal' 
millimetre line emission towards the two 
oxygen-rich, low mass loss rate AGB stars  \object{R Dor} and  \object{L$^2$ Pup} are presented. In both cases the emission is resolved with an overall spherical symmetry. Detailed radiative transfer modelling of the SiO line emission has been performed, and the comparison between observations and models are conducted in the visibility plane, maximizing the sensitivity. The excitation analysis suggests that the abundance of SiO is as high as $4\times 10^{-5}$ in the inner part of the wind, close to the predicted values from stellar atmosphere models.  Beyond a radius of $\approx 1\times 10^{15}$ cm the SiO abundance is significantly  lower, about  $3\times 10^{-6}$, until it decreases strongly at a radius of about   $3\times 10^{15}$ cm.
This is consistent with a scenario where SiO first freezes out onto dust grains, and then
eventually becomes photodissociated by the interstellar UV-radiation field.
In these low expansion velocity sources  the turbulent broadening of the lines plays an important role in the line formation. Micro-turbulent velocity widths in the range $1.1-1.5$ km\,s$^{-1}$ result in a very good reproduction of the observed line shapes even if the gas expansion velocity
is kept constant. This, combined with the fact that the SiO and CO lines are well fitted using the same gas expansion velocity (to within $5-10$\%), suggest that the envelope 
acceleration occurs close to the stellar photosphere, within $\lesssim 20-30$ stellar radii. 


   }
   \maketitle
%

\section{Introduction}
The intense winds that  low to intermediate mass stars develop in their final evolutionary stage, as they ascend the asymptotic giant branch (AGB), return enriched stellar material to the interstellar medium. In addition to significantly contributing to the chemical evolution of galaxies, the mass loss will dictate the time scale for the future evolution of the star towards the planetary nebula phase. Considering its importance little is known with
certainty about the mechanism(s) behind the mass loss. Partly this is due to a lack of observational constraints, in particular close to the stellar photosphere 
where the wind is accelerated, and partly due to the complexity of the
physical/chemical processes involved. Among other things, the formation
of dust grains is thought to play an important role in these radiatively driven winds.
Most of the information on this region comes from
observations of infrared ro-vibrational molecular lines in absorption, mainly towards
the carbon star \object{IRC+10216} \citep[e.g.,][]{Keady93,Winters00a}, infrared
continuum emission from the circumstellar dust \citep[e.g.,][]{Danchi94}, and SiO maser 
radio line emission from excited vibrational states \citep[e.g.,][]{Cotton04}.

Information can also be gained from radio observations of suitable circumstellar molecular species
towards stars having low to intermediate mass loss rates.
A major survey of CO radio line emission from circumstellar envelopes (CSEs) around oxygen-rich AGB stars of different variability types were 
done by \citet{Kerschbaum99} and \citet{Olofsson02}. These data were modelled  in detail to derive stellar mass loss rates and terminal gas expansion velocities \citep{Olofsson02}. 
Subsequently, a survey of `thermal' SiO radio line emission, meaning emission from the ground vibrational state rotational lines which are normally not (strongly)
masering, was done and the data were interpreted using
a detailed numerical radiative transfer modelling presented in \citet{Delgado03b}.  

An immediate conclusion from these large surveys is that the SiO and CO radio line profiles are different from each other.  
The SiO line profiles are narrower in the sense that the main
fraction of the emission comes from a velocity range
smaller by about $10-20$\% than twice the expansion
velocity determined from the CO data.  On the other hand, the SiO line
profiles have weak wings, such that the total velocity width of its
emission is very similar to that of the CO emission. Furthermore, it appears that the SiO line profiles change character with the mass loss rate, at low mass loss rates they are narrow with weak extended wings, while at high mass loss rates they become distinctly triangular.  These
`peculiar' SiO line profiles have been interpreted as being due to the influence
of gas acceleration in the region which produces most of the SiO line
emission \citep{Bujarrabal86}. However, as illustrated in \citet{Delgado03b}, 
the SiO lines are usually strongly self-absorbed also for low mass loss rate objects 
and this produces narrower lines. There is some controversy in the literature over the scale length of the acceleration region. In the only previously published works where thermal SiO emission have been observed towards AGB stars using interferometry, \citet{Sahai93} find no need for the slowly varying velocity fields introduced by \citet{Lucas92}.

Other conclusions come from the detailed radiative transfer modelling of the SiO line
data. This is in many respects a more difficult
enterprise than the CO line modelling.  The SiO line emission predominantly 
comes from a region closer
to the star than does the CO line emission, and this is a region where
we have fewer observational constraints.  The SiO excitation is also
normally far from thermal equilibrium with the gas kinetic
temperature, and radiative excitation plays a major role.  Finally,
there exists no detailed chemical model for calculating the radial SiO
abundance distribution. \citet{Delgado03b}  adopted the assumption that 
the gas-phase SiO abundance stays high only very close to the star, since
further out the SiO molecules are adsorbed onto the grains.  Beyond
this the abundance stays low until the molecules are eventually
dissociated by the interstellar UV radiation.  This photodissociation 
radius, which is 
crucial to the modelling, was estimated
using both SiO multi-line modelling and existing interferometer data
\citep{Lucas92, Sahai93}, but only for a few sources. The result of
the radiative transfer modelling is a circumstellar SiO abundance that
is roughly the same as that obtained from stellar atmosphere equilibrium
chemistry for low mass loss rate objects, and which declines with mass
loss rate reaching an abundance about two orders of magnitude lower for high 
mass loss rate objects.

Thus, there are strong indications that `thermal' SiO radio line emission is a useful probe of the formation and
evolution of dust grains in a CSE, as well as of its dynamics.
Hence, circumstellar SiO line emission potentially carries information on the properties of the region where the mass loss of AGB stars is initiated.

Presented here are results from high spatial resolution imaging of the two M-type semiregular AGB stars \object{R Dor} and  \object{L$^2$ Pup}.
Both sources studied have low mass loss rates ($\dot{M}\lesssim 1\times 10^{-7}$ M$_\odot$\,yr$^{-1}$) and low terminal 
velocities of their winds ($\lesssim6$ km\,s$^{-1}$).
It has been suggested from hydrodynamical calculations \citep{Winters00b,Winters02,Winters03} that 
the main driving mechanism behind these tenuous winds is stellar pulsation and that dust plays only a secondary role. 

The observations, performed with the Australia Telescope Compact Array\footnote{The Australia Telescope is funded by the Commonwealth of Australia for
operation as a National Facility managed by CSIRO.} (ATCA), are presented in Sect.~2.
ATCA is an array of six 22 m dishes
operating from $1.4-26$ GHz and with an upgrade to $85-105$ GHz operation in
progress.  Its location makes it a unique instrument to study molecular-line
sources in the southern hemisphere at high angular resolution, and we have
recently used it to conduct a study of circumstellar HCN emission from the
carbon star R Scl \citep{Wong04}. The analysis in Sect.~3 and comparison with envelope models in Sect.~4 are carried out in the $uv$-plane in order to maximize the sensitivity and resolution of the data. The modelling is followed by a discussion in Sect.~5 and conclusions are presented in Sect.~6.
 

%
\begin{figure}
\centerline{\includegraphics[width=7cm]{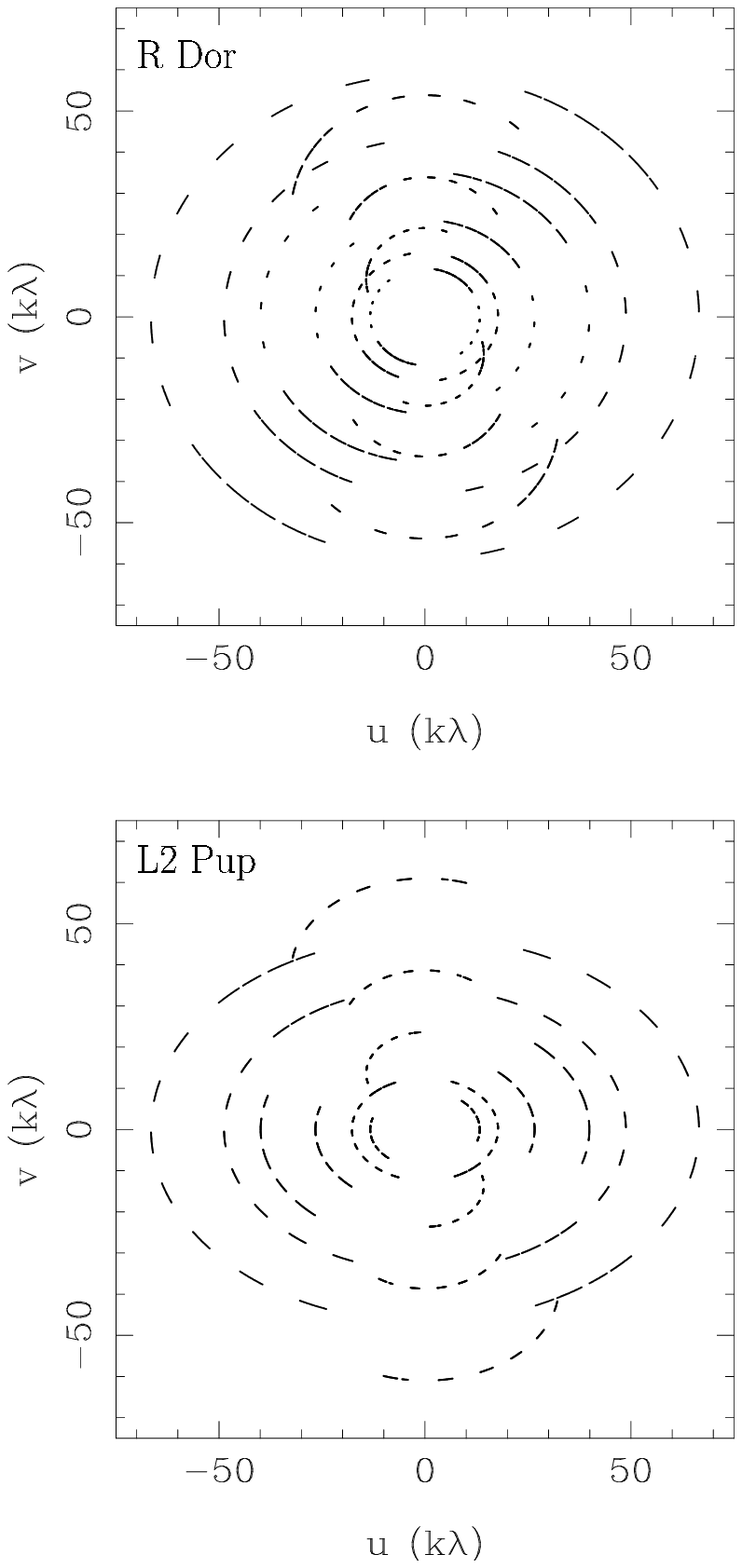}}
  \caption{Coverage of the visibility plane  for \object{R Dor} and \object{L$^2$ Pup} obtained after combining the three ATCA configurations used (EW367, 750B, and H214).} 
  \label{uvcov}
\end{figure}

\section{Observations and data reduction}
\label{sec:obs}

At the time of the observations, the ATCA had three antennas of 22~m
diameter equipped with dual polarisation 3-mm receivers covering the
bands $84.9-87.3$ and $88.5-91.3$ GHz.  We observed 
\object{R~Dor} and \object{L$^2$ Pup} between
August and October 2003 in three different array configurations: EW367
(baselines of 45, 90, and 135 m), 750B (60, 165, and 225 m), and H214 (75,
135, and 210 m).  EW367 and 750B provided east-west baselines whereas
H214 provided north-south baselines.  The observations in each
configuration occurred on single nights. 
The combined visibility plane coverage is shown in Fig.~\ref{uvcov}.

All observations were conducted in clear to partly cloudy weather, with
above-atmosphere single-sideband system temperatures of $T_{\rm sys}^*
\approx 350$--500~K near the zenith.  The correlator was configured to
receive both linear polarisations in two frequency windows: a
narrowband (spectral line) window centred on the `thermal' SiO $v=0,
J=2\rightarrow1$ 
line at 86.847~GHz with 128 channels across 32 MHz, and a wideband
window centred  on the `maser' SiO $v=1, J=2\rightarrow1$
line at 86.243 GHz with 32 channels across 128
MHz. The pointing and phase centre was at
$\alpha_{2000} = 04^{\rm h} 36^{\rm m} 45\fs67$, $\delta_{2000} =
-62\degr 04\arcmin 37\farcs9$ 
for \object{R Dor} and at $\alpha_{2000} = 07^{\rm h}
13^{\rm m} 32\fs31$, $\delta_{2000} = -44\degr 38\arcmin 24\farcs1$
for \object{L$^2$ Pup}.  The field of view is determined by the ATCA
primary beam which has a {\em FWHM} of about $36\arcsec$ at 86~GHz.

As an initial step in the reduction of the 750B and H214 data, an
elevation-dependent gain curve was applied based on observations of an
SiO maser taken on 2003 Sep 4.  No such gain curve was available for the
earlier EW367 observation.  Gain calibration as a function of time was then
performed by assuming the SiO maser emission comes from a point source
located at the phase centre.  Since the maser emission is known to
emanate from a region only a few AU in radius \citep[e.g.,][]{Cotton04}, this is
a reasonable assumption for our resolution of a few arcseconds (1 AU = 20
milliarcseconds at the distance of our nearest source, R Dor).  To
derive the amplitude gains we also assumed that the integrated SiO
maser flux was constant throughout each observation. 
We found that the variation in the gain
was generally dominated by an uncorrected gain-elevation dependence,
limiting any intrinsic flux variation to $\lesssim10$\%.

\begin{figure*}
\centerline{\includegraphics[height=14cm,angle=-90]{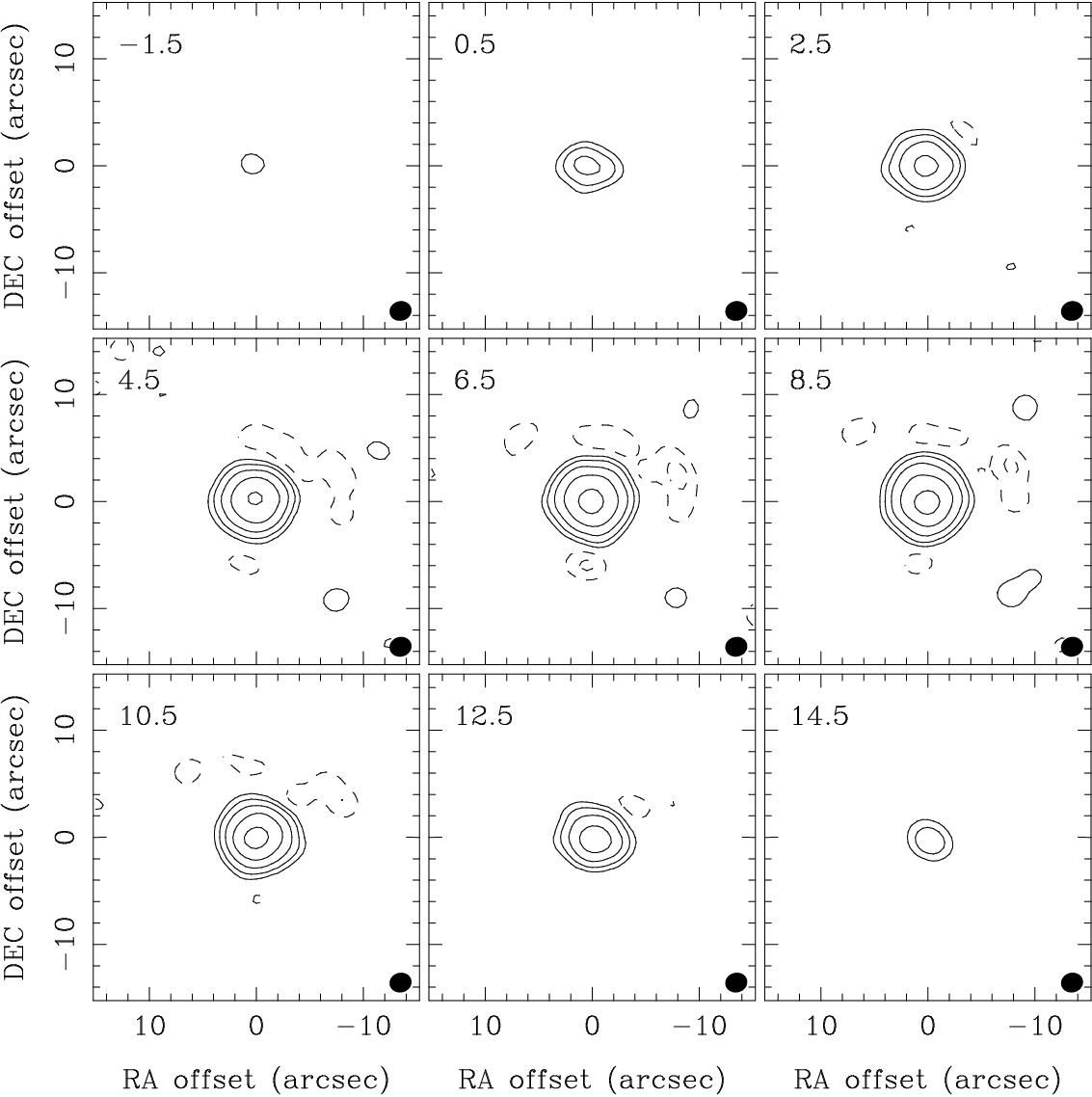}}
  \caption{Velocity channel maps of SiO $v=0, J=2\rightarrow1$ emission from \object{R Dor}.  
The contour levels are $0.03n$ Jy~beam$^{-1}$, where $n=-9,-3,3,9,27,81,243$ (negative values have dashed contours), and the beam size
is $2\farcs6\times2\farcs0$ with a position angle of $-56\degr$ as indicated in the lower right hand corner of each panel.  The velocity channels (given in the
LSR frame and indicated in the upper left corner) have been binned to 2 km~s$^{-1}$. The systemic velocity is 7 km\,s$^{-1}$ as determined from CO observations.
Offsets in position are relative to $\alpha_{2000} = 04^{\rm h} 36^{\rm m} 45\fs67$, $\delta_{2000} =
-62\degr 04\arcmin 37\farcs9$.} 
  \label{rdor_cm}
\end{figure*}

The amplitude and phase gains were transferred from the maser to the
spectral line window after applying scaling factors determined from a
half-hour integration on the bandpass calibrator (3C279 or PKS
B1921-293), which was also used to determine the channel-dependent
gains.  In the case of \object{L$^2$ Pup}, the transfer of the phase gains could
not be performed successfully because of an inconsistency in the
frequency setup between the source and bandpass calibrator;
consequently we have applied phase self-calibration to the thermal SiO
emission from this source assuming that it is circularly symmetric.
For an interferometer the astrometry is determined by the quality of the
phase calibration.  For \object{R Dor} the phase errors were $\sim 5\degr$ and 
one would expect the position error to be about $0.1\arcsec$
for a 100 m baseline. 

Uranus was used to set the flux scale, assuming a uniform disk with
brightness temperature of 134~K.  However, for the EW367 array, where
the Uranus observation occurred in poor weather, the flux scale was
tied to that of the 750B observation by assuming the SiO maser fluxes
from both sources were unchanged over the 17-day gap between the
observations.  Assuming that any actual flux variations for the two
sources were uncorrelated, the derived gains suggest an uncertainty
in the flux scale for this configuration of $\approx 30$\%.  We also
adjusted the antenna pointing once an hour on the SiO masers; typical
pointing shifts were $\approx 5\arcsec$.

All data processing was conducted using the MIRIAD package
\citep{Sault95}.  The calibrated visibilities were Fourier transformed
using uniform weighting and a $0\farcs5$ pixel size with a channel
spacing of 1 km\,s$^{-1}$ (roughly the effective velocity resolution
given the original channel spacing of 0.25 MHz $\approx 0.8$
km\,s$^{-1}$).  The maps were then CLEANed \citep{Hoegbom74} down to a
2$\sigma$ level over the inner 20\arcsec$\times$20\arcsec.

The actual analysis and comparison with the model will be carried out
in the $uv$-plane to maximize the sensitivity and resolution of the
data.  Thus we expect to obtain usable information on scales as low as
$1\arcsec$, corresponding to the longest baselines.

\section{Observational results}
\label{sec:obsres}

\subsection{R Dor}
The SiO $v=0, J=2\rightarrow1$ velocity channel maps of  \object{R Dor},
binned into 2 km\,s$^{-1}$ channels,
are shown in Fig.~\ref{rdor_cm}. The synthesised beam is $2\farcs6\times2\farcs0$ with a position angle of $-56\degr$ and 
the RMS noise in the 1 km\,s$^{-1}$ velocity channel maps is $\sigma = 30$~mJy\,beam$^{-1}$, as estimated from regions free of source emission.
The signal-to-noise ratio is high with the peak emission reaching a level of about $300\sigma$.  

\begin{figure}
\centerline{\includegraphics[width=8cm]{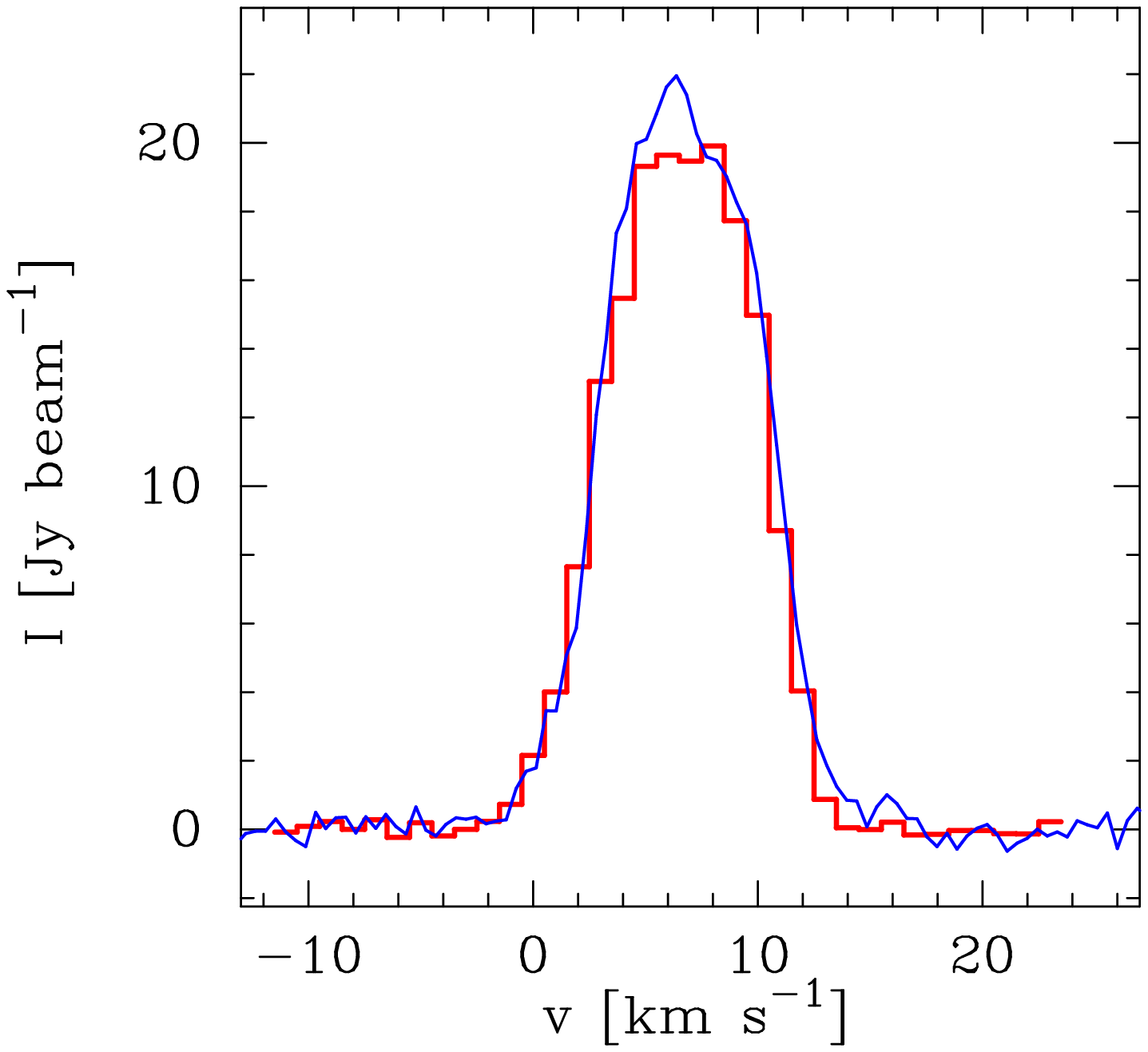}}
  \caption{SiO $v=0, J=2\rightarrow1$ spectra for  \object{R~Dor}. 
 The solid line is the SEST single dish observation \citep{Delgado03b}, whereas the histogram shows the spectrum at the phase centre  derived from the ATCA data by restoring with the SEST beam of $57\arcsec$.} 
  \label{rdorsest}
\end{figure}

The velocity channel maps suggest that the SiO $v=0, J=2\rightarrow1$ emission towards  
\object{R Dor} is moderately
resolved, a result which is confirmed by the visibility analysis (see below). 
The brightness distribution appears to have an overall circular symmetry.
From fitting two-dimensional Gaussian brightness distributions 
to the individual channel maps, it is concluded that 
no systematic variations in position with velocity are present. The individual offsets are smaller than $0\farcs1$, within the pointing uncertainty (Sect.~\ref{sec:obs}), confirming the adopted source position.  
Across the peak of the line profile, from $6-8$~km\,s$^{-1}$ (LSR), 
the average values of the major and minor axes are 3\farcs3 and 3\farcs7 ({\em FWHM}), respectively.
The beam has not been deconvolved from these values.

\begin{figure}
\centerline{\includegraphics[width=8cm]{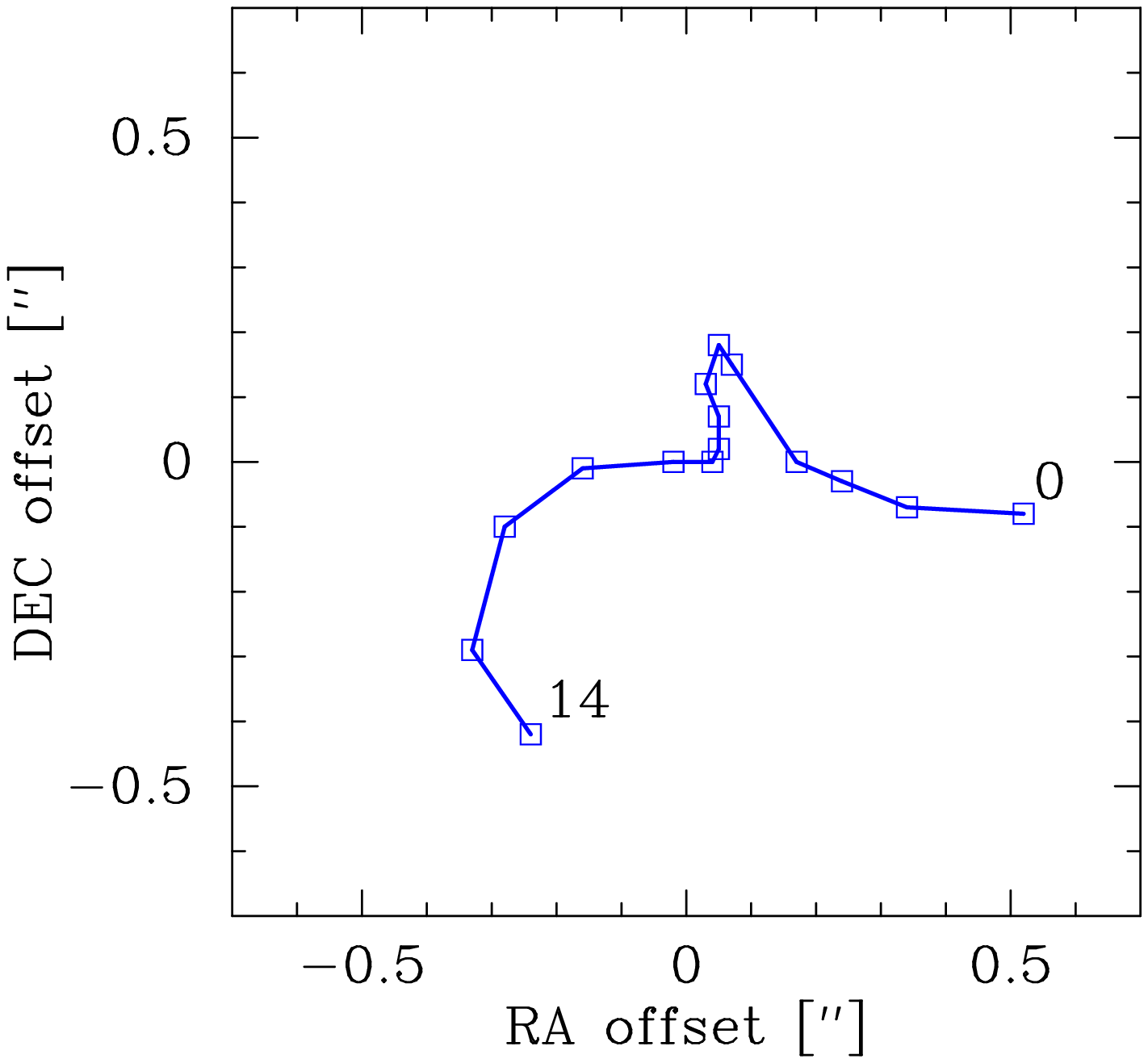}}
  \caption{Offset from the phase centre of the velocity channels as determined from fitting circular Gaussians to the \object{R~Dor} data in the $uv$-plane. Indicated are the velocities of the two extreme channels (0 and 14 km\,s$^{-1}$) for which reliable fits could be obtained.} 
  \label{rdor:offset}
\end{figure}
\begin{figure}
\centerline{\includegraphics[width=8cm]{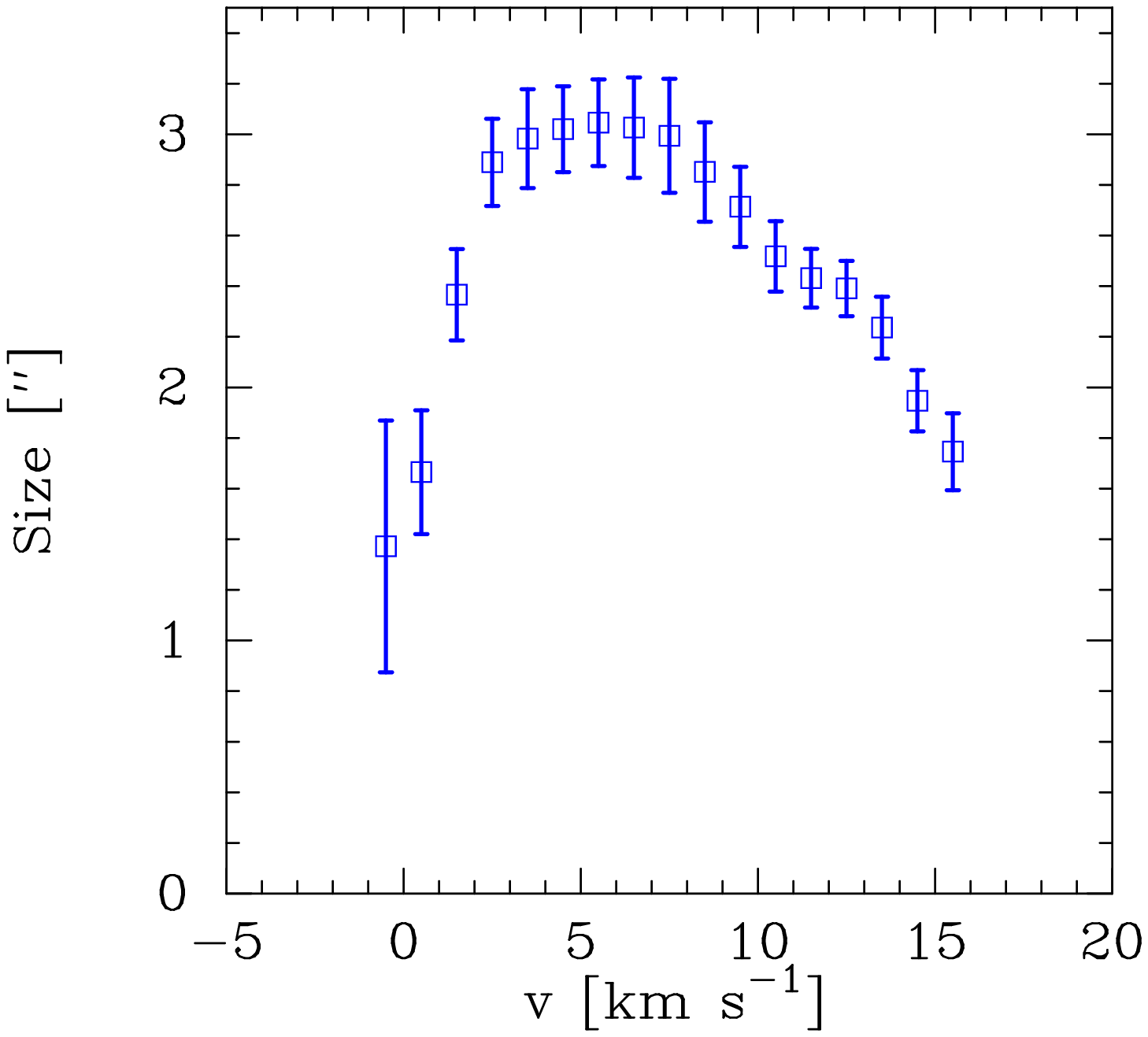}}
  \caption{Size ({\em{FWHM}}) of the observed SiO emission toward  
  \object{R~Dor} as function of velocity channel (given in LSR frame), averaged in 1 km\,s$^{-1}$ bins, as estimated from Gaussian fits to the azimuthally averaged visibility data.} 
  \label{rdoruvfit}
\end{figure}

As interferometers lacks sensitivity to large scale emission  it is of interest to find out whether part of the emission is resolved out in the ATCA data. In Fig.~\ref{rdorsest} the ATCA
SiO data has been restored with a $57\arcsec$ circular beam (solid line; this represents the beam of the Swedish-ESO Submillimetre Telescope, SEST). A comparison with the equivalent SEST spectrum from 
\citet{Delgado03b} (histogram) shows a striking resemblance both with regard to the peak flux ($\approx 20$ Jy) and the overall line shape. This strongly suggests that all the flux is recovered by the ATCA interferometer, at least within the $57\arcsec$ SEST beam, and
that virtually all of the SiO emission comes from the central component. 
The SEST spectrum has been converted from main beam brightness temperature scale to Jy using 
\begin{equation}
\label{K2Jy}
S = \eta_{\rm mb}  \Gamma^{-1} T_{\mathrm{mb}}, 
\end{equation}
where the main-beam efficiency $\eta_{\rm mb}=0.75$ and the sensitivity $\Gamma^{-1}=25$ Jy\,K$^{-1}$. The telescope parameters are taken from the SEST homepage\footnote{\tt{www.ls.eso.org/lasilla/Telescopes/SEST/SEST.html}}.

Assuming that the emission has an overall spherical symmetry, circularly symmetric Gaussians have been fitted fitted to the visibilities. Given the higher sensitivity in the $uv$-plane there appears to be a systematic trend with velocity as shown in Fig.~\ref{rdor:offset}. The largest offsets are measured at the extreme velocities, i.e., the blue- and redshifted emissions are displaced from each other. \citet{Lindqvist00}, when analysing several molecular line emissions from the high mass loss rate carbon star \object{CIT 6}, also noted that there was a common trend
in position with velocity. A reasonable explanation for this behaviour is that the CSE is asymmetric to some degree, perhaps a bipolar outflow,
but the present angular resolution is not enough to make this apparent in the channel maps,
Fig.~\ref{rdor_cm}.

The resulting {\em FWHM} of Gaussians fitted 
to the azimuthally averaged visibilities are plotted in Fig.~\ref{rdoruvfit}. The overall variation in size with line-of-sight velocity is as expected for a well resolved expanding envelope, where gas moving orthogonal to the line of sight subtends a larger solid angle than radially moving gas near the extreme velocities.
The largest angular extent of the emission is $3\farcs0\pm0\farcs2$ in the velocity range $6-8$ km~s$^{-1}$ (LSR). This corresponds to a radial size of $1.0\times 10^{15}$ cm at the distance of 45 pc. We note that a single Gaussian provides a reasonable fit to the visibilities on all baselines except those close to the systemic velocity at 7 km\,s$^{-1}$ (LSR). Introducing an offset in the flux scale, representing an unresolved additional component, significantly improves the fit at those velocities. In Sect.~\ref{RDor:mod} it is shown that the visiblilities obtained for this source are best fitted using two components: one compact with high a SiO abundance and one more extended with a significantly lower abundance.

\begin{figure*}
\centerline{\includegraphics[height=14cm,angle=-90]{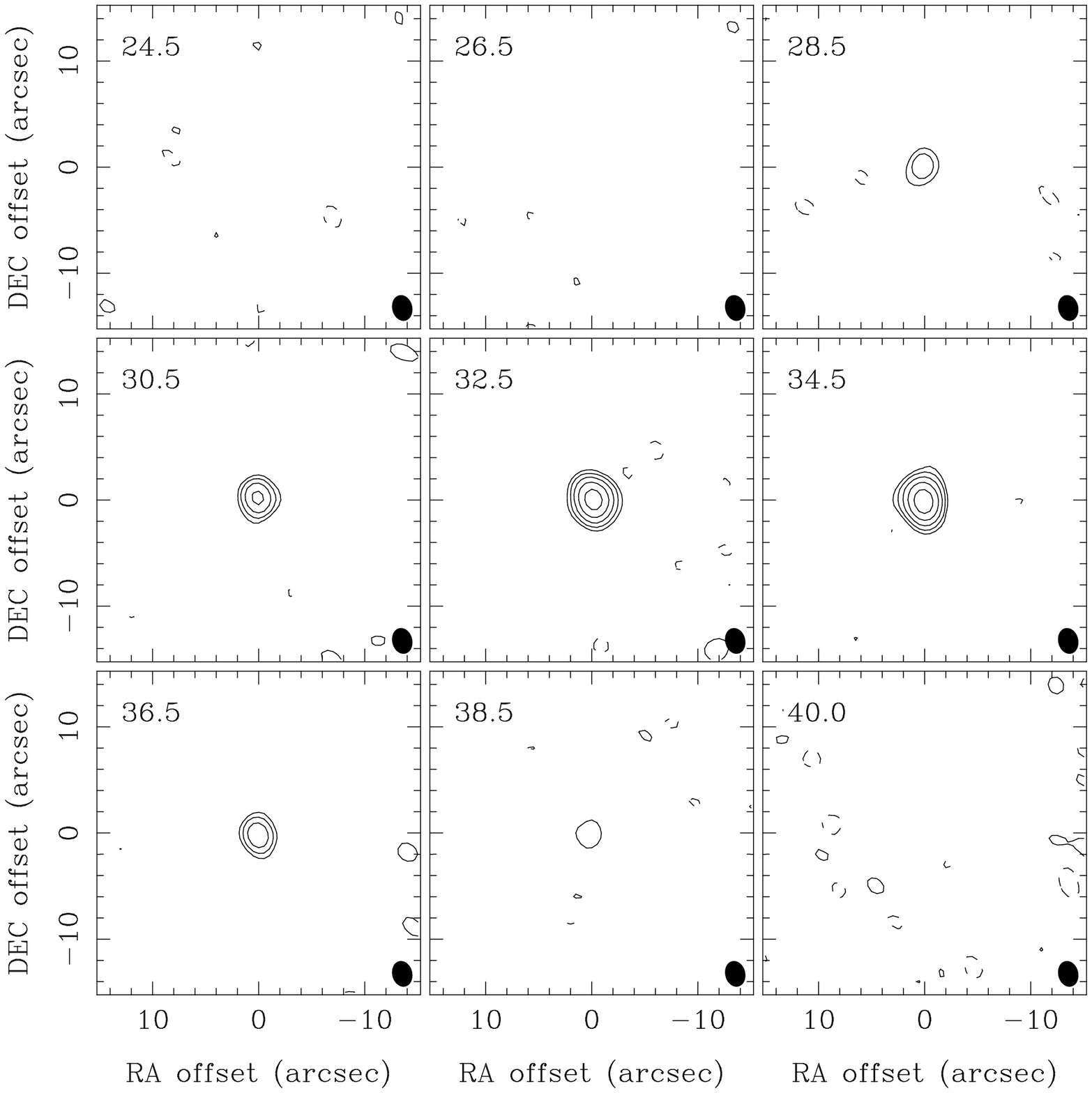}}
  \caption{Velocity channel maps of SiO $v=0, J=2\rightarrow1$ emission from \object{L$^2$ Pup}.  
The contour levels are $0.04n$ Jy~beam$^{-1}$, where $n=-2, 2, 4, 8, 16, 32$ (negative values have dashed contours), and the beam size
is $2\farcs5\times1\farcs9$ with a position angle of $14\degr$ as indicated in the lower right hand corner of each panel..  The velocity channels (given in the
LSR frame and indicated in the upper left corner) have been binned to 2 km~s$^{-1}$. The systemic velocity is 33 km\,s$^{-1}$ as determined from CO observations. Offsets in position are relative to $\alpha_{2000} = 07^{\rm h}
13^{\rm m} 32\fs31$, $\delta_{2000} = -44\degr 38\arcmin 24\farcs1$.} 
  \label{l2pup_cm}
\end{figure*}

\subsection{L$^2$ Pup}
The SiO $v=0, J=2\rightarrow1$ velocity channel maps of \object{L$^2$ Pup},
binned into 2 km\,s$^{-1}$ channels,
are shown in Fig.~\ref{l2pup_cm}. The synthesised beam is $2\farcs5\times1\farcs9$ with a position angle of $14\degr$ and 
the RMS noise in the 1 km\,s$^{-1}$ velocity channel maps is $\sigma = 40$~mJy\,beam$^{-1}$,
as estimated from regions free of source emission.
The signal-to-noise ratio is high with the peak emission reaching
a level of about 60$\sigma$.  

As expected from the single-dish modelling performed by \citet{Delgado03b} \object{L$^2$ Pup} has a smaller envelope than \object{R Dor}. Since we had to self-calibrate the data in phase for this object, we have assumed that the source distribution is symmetric and centred on the stellar position. Across the peak of the line profile, from $33$ to $35$ km\,s$^{-1}$ (LSR), the average values of the major and minor axes are 2\farcs9 and 2\farcs2 
({\em FWHM}), respectively (note that the beam has not been deconvolved from these values).

\begin{figure}
\centerline{\includegraphics[width=8cm]{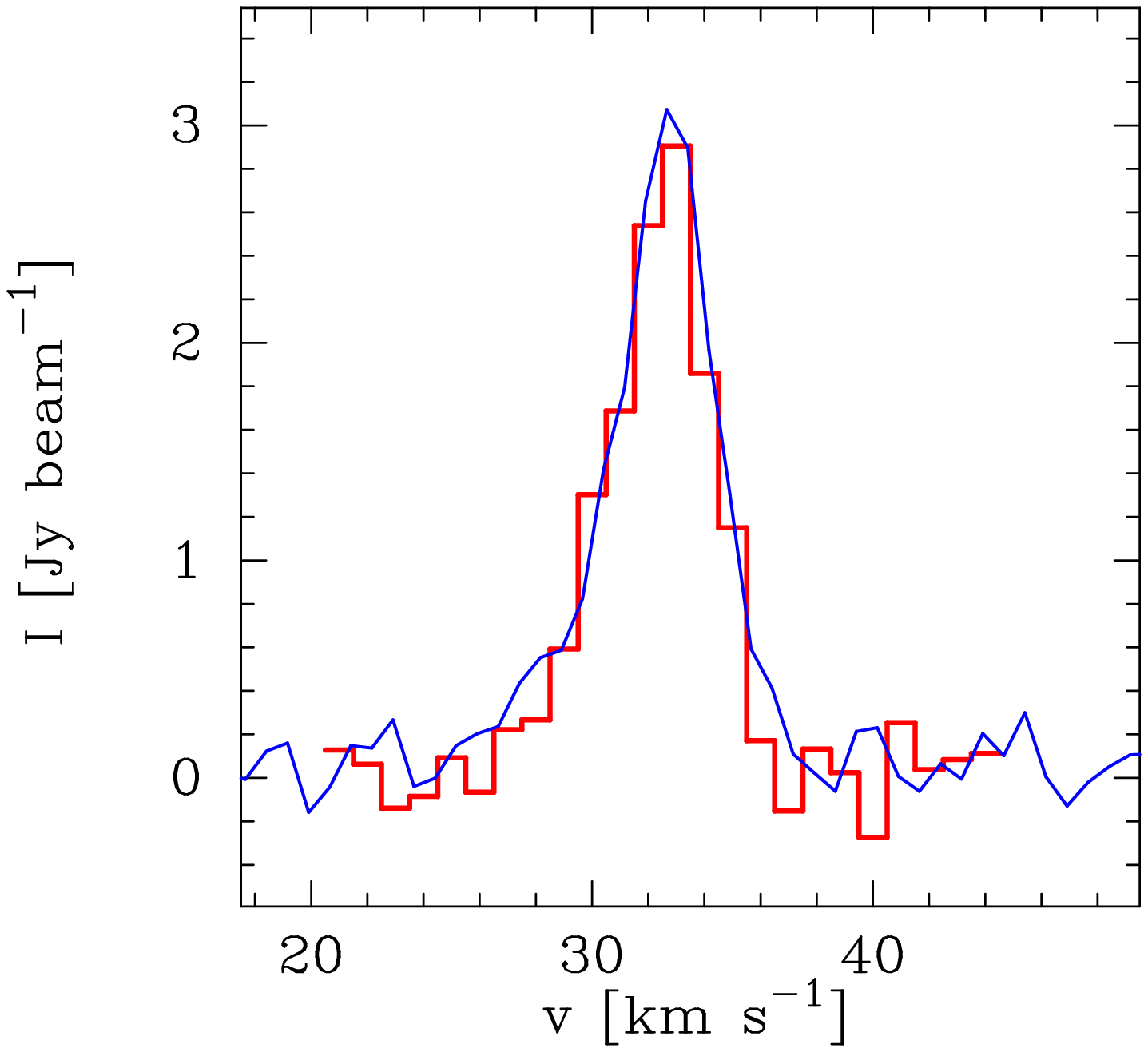}}
  \caption{SiO $v=0, J=2\rightarrow1$ spectra for  \object{L$^2$ Pup}. 
 The solid line is the SEST single dish observation \citep{Delgado03b}, whereas the histogram shows the spectrum at the phase centre  derived from the ATCA data by restoring with the SEST beam of $57\arcsec$.} 
  \label{l2pupsest}
\end{figure}

In Fig.~\ref{l2pupsest} the ATCA SiO data for \object{L$^2$ Pup} has been restored with a $57\arcsec$ circular beam (solid line), representing the SEST beam. A comparison with the equivalent SEST spectrum of \citet{Delgado03b}  (histogram; converted from main beam brightness temperature scale to Jy using Eq.~\ref{K2Jy}) shows a striking resemblance both 
with regard to the peak flux ($\approx 3$ Jy) and the overall line shape. As for \object{R Dor} this strongly suggests that all the flux is recovered by the ATCA interferometer, and that all of the SiO emission comes from the central compact component.

\begin{figure}
\centerline{\includegraphics[width=8cm]{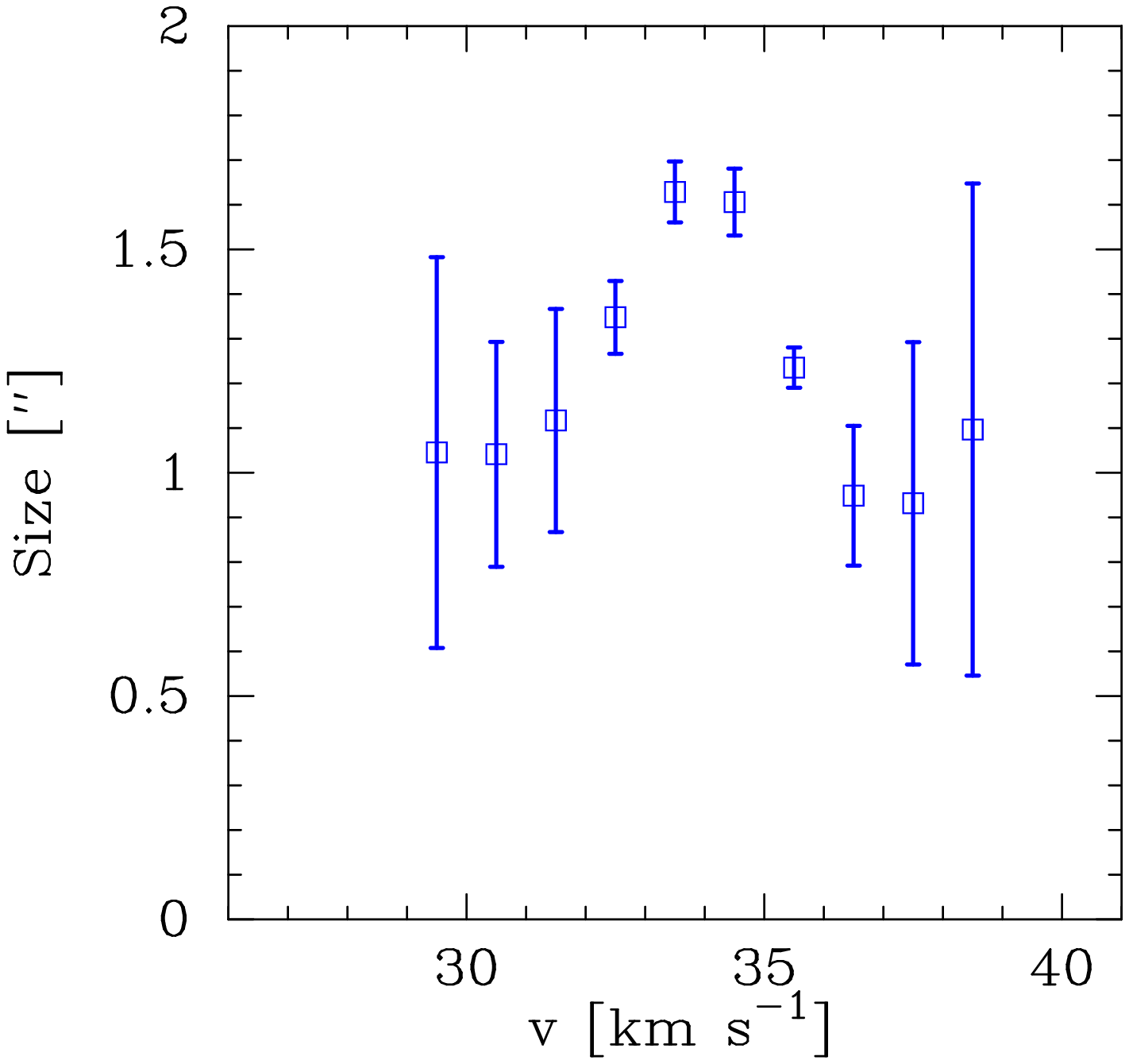}}
\caption{Size ({\em{FWHM}}) of the observed SiO emission toward  
  \object{L$^2$~Pup} as function of velocity channel (given in LSR frame), averaged in 1 km\,s$^{-1}$ bins, as estimated from Gaussian fits to the azimuthally averaged visibility data.} 
  \label{l2pupuvfit}
\end{figure}

The result of fitting Gaussians to the azimuthally averaged visibilities is presented in Fig.~\ref{l2pupuvfit}. The variation in size is also here consistent with a resolved expanding envelope.
The largest angular extent of the emission is $1\farcs6\pm0\farcs1$ in the velocity range $33-35$ km\,s$^{-1}$ (LSR). This corresponds to a radial size of $1.0\times 10^{15}$ cm at the distance of 85 pc.
Contrary to \object{R Dor}, a Gaussian gives a reasonably good description of the visibilities on all baselines and for all velocity channels.


\section{Radiative transfer modelling}
A detailed non-LTE radiative transfer code, based on the Monte Carlo method
\citep{Bernes79}, is used to perform the excitation analysis and model the observed  
circumstellar line emission.  
The code is described in detail in \citet{Schoeier01} and has been
benchmarked, to high accuracy, against a wide variety of molecular line radiative
transfer codes in \citet{Zadelhoff02}.  

\subsection{The physical structure of the CSE}
The CSEs are assumed to be spherically symmetric and to expand at constant velocity, and they are produced by  constant mass loss rates.
Both \object{R~Dor} and  \object{L$^2$ Pup} were included in the large survey of semiregular M-type AGB-variables by \citet{Kerschbaum99}. Their overall wind properties such as mass loss rate, terminal expansion velocity and temperature structure as constrained by single-dish CO millimetre line emission were presented in \citet{Olofsson02}, where also a more detailed description of the circumstellar model is given. 

\citet{Olofsson02} noted that the CO lines for \object{L$^2$ Pup} are better reproduced when adopting a higher value of 1.0 km\,s$^{-1}$ for the turbulent velocity than the nominal value of 0.5 km\,s$^{-1}$. Further support for a larger turbulent velocity are the SiO line wings that are commonly observed in the low expansion velocity sources such as \object{R~Dor} and  \object{L$^2$ Pup} \citep{Delgado03b}. Therefore, in the present analysis the turbulent
velocity width is used as a free parameter.

\citet{Olofsson02} calculated the kinetic temperature distribution including, e.g., CO line cooling. Here we add cooling by H$_2$O since this may be of importance
in the region where the SiO line emission is produced \citep{Delgado03b}. 
The importance of H$_2$O as a coolant in the inner wind was also noted by \citet{Zubko00} in their study of \object{W Hya}. In order to be consistent we have therefore produced new circumstellar
models for also the CO line emission. 
The inclusion of H$_2$O line cooling lowers the temperature significantly (from $15-40$\%) in the inner part ($r\lesssim 1\times 10^{15}$ cm) of the envelope, but has only a little effect on the mass loss rate derived from CO observations. More details on how the  H$_2$O line cooling is treated, and its effect on CO and SiO line intensities, are given in \citet{Olofsson02} and \citet{Delgado03b}. As an example, we note that excluding H$_2$O line cooling increases the predicted SiO line intensities in our best fit model for \object{R Dor} (presented in Sect.~\ref{RDor:mod}) by $10-15$\% for the $J=5\rightarrow4$ and $J=6\rightarrow5$  SEST observations as well as the $J=2\rightarrow1$ interferometric ATCA observations. These are the data most sensitive to the physical and chemical conditions prevailing in the inner wind. For comparison, the predicted SEST $J=2\rightarrow1$ and $J=3\rightarrow2$ line intensities only increase by a few \%.

We find that the molecular line emission is better reproduced (in particular that of SiO) using a turbulent width of 1.1 km\,s$^{-1}$ for \object{L$^2$ Pup} and 1.5 km\,s$^{-1}$ for \object{R~Dor}. Adopting a larger turbulent width means that the expansion velocity needs to be lowered somewhat in order to still provide good fits to the line widths. For  for \object{L$^2$ Pup} an expansion velocity of 2.1 km\,s$^{-1}$ was found and for \object{R~Dor} 5.3 km\,s$^{-1}$. These values are about 20\% lower than those presented in \citet{Delgado03b} where a much lower microturbulent velocity of  0.5 km\,s$^{-1}$ 
was adopted. Changing the prescription of the velocity field, and in particular the turbulent velocity, changes the excitation conditions and somewhat different mass loss rates are obtained. For \object{L$^2$ Pup} a mass loss rate of $2.7\times 10^{-8}$ M$_{\sun}$\,yr$^{-1}$ is found to provide a very good fit using an $h$-parameter of 0.05 typical of low mass loss rate objects. The $h$-parameter contain much of the properties of the dust, such as size and density of individual grains and the dust-to-gas ratio, and enters in the heating term in the energy balance equation which is solved self-consistently in the excitation analysis (see Sch\"oier \& Olofsson\ 2001\nocite{Schoeier01} for a more elaborate explanation of the $h$-parameter). For \object{R~Dor} a mass loss rate of $1.2\times 10^{-7}$ M$_{\sun}$\,yr$^{-1}$ is obtained with an $h$-parameter of 0.05. 

\begin{table}
\caption[]{Source parameters obtained from modelling CO millimetre line emission.}
  \label{parameters}
$$
\begin{array}{p{0.15\linewidth}ccccccc}
\hline\hline
\noalign{\smallskip}
Source & D\,^{\mathrm a} & L_{\star}\, ^{\mathrm a} & T_{\star}\, ^{\mathrm a} & \dot{M} & h &v_{\infty} & v_{\mathrm t} \\
&            [\mathrm{pc}] & [\mathrm{L}_{\sun}] & [\mathrm{K}] & [\mathrm{M}_{\sun} \mathrm{yr}^{-1}] & &  \multicolumn{2}{c}{\mathrm{|\mathrm{km\ s}^{-1}]}}\\
\noalign{\smallskip}
\hline
\noalign{\smallskip}
R Dor   & 45 & 4000 & 2090 & 1.2\times 10^{-7} & 0.05 & 5.3 & 1.5 \\
L$^2$ Pup & 85 & 4000 & 2690 & 2.7\times 10^{-8} & 0.05 & 2.1 & 1.1\\
\hline
\end{array}
$$
\noindent
$^{\mathrm a}$ From \citet{Olofsson02}
\end{table}
\begin{figure}
\centerline{\includegraphics[width=8cm]{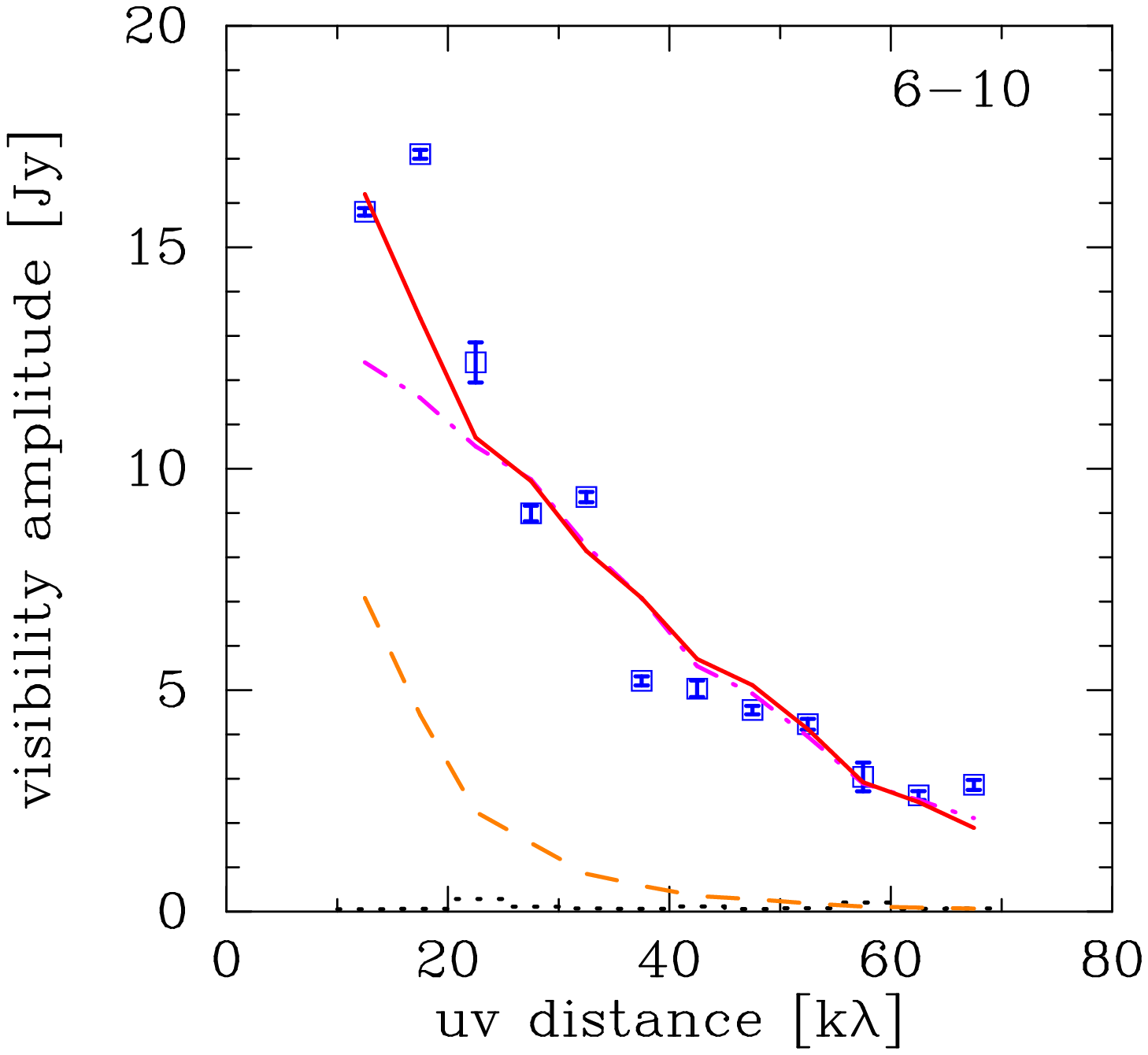}}
  \caption{Azimuthally averaged visibility amplitudes, in 5 k$\lambda$ bins,
across the peak of the line profile ($6-10$ km\,s$^{-1}$ in the LSR frame) for \object{R Dor}. The error bars indicate the 2$\sigma$ statistical error and does not contain the calibration uncertainty of about 15\%. The solid line indicates the best fit model ($X_0=4\times 10^{-5}$; $r_{\mathrm c}=1.2\times 10^{15}$ cm;  $X_{\mathrm D}=3\times 10^{-6}$;  $r_{\mathrm e}=3.3\times 10^{15}$ cm), the dashed line the best fit model from \citet{Delgado03b} ($X_0=5\times 10^{-6}$; $r_{\mathrm e}=3.3\times 10^{15}$ cm), and the dashed-dotted line the best fit model without freeze-out ($X_0=5.5\times 10^{-5}$; $r_{\mathrm e}=1.0\times 10^{15}$ cm). The dotted histogram indicates the zero-expectation level, i.e., when no source emission is present.} 
  \label{rdor_velvis}
\end{figure}
\begin{figure*}
\centerline{\includegraphics[width=17cm]{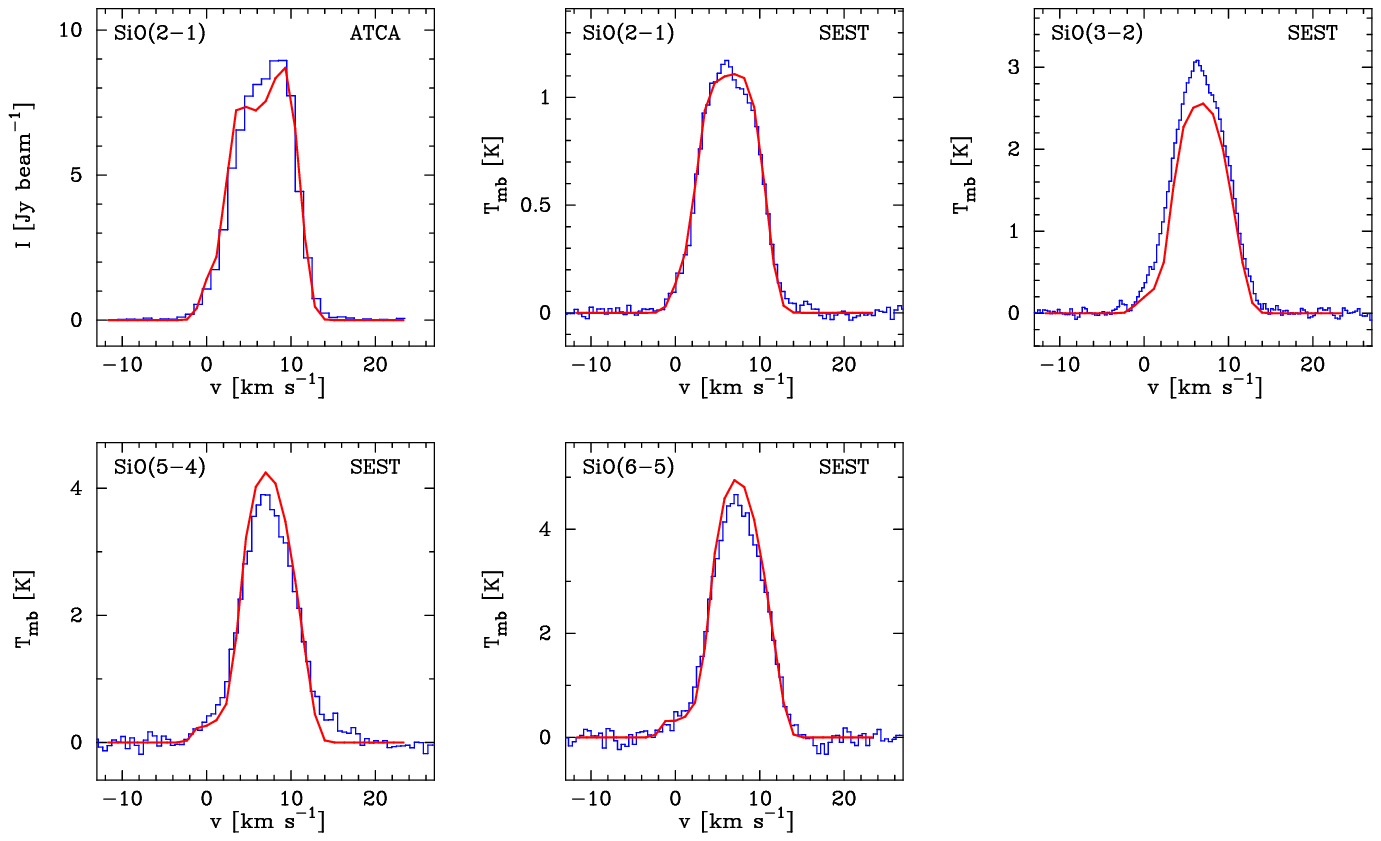}}
  \caption{Best fit model (solid line) overlayed on observations (histograms) in the form of single-dish SEST data \citep{Delgado03b} as well as the spectrum at the centre pixel in the cleaned ATCA map for \object{R Dor}. The velocity scale is given in the LSR frame. The calibration uncertainty in the single-dish data is about $\pm 20$\%.} 
  \label{rdorspec}
\end{figure*}

The resulting envelope parameters
are summarized in Table~\ref{parameters}. The distances adopted are from \citet{Olofsson02} and are based on the assumption that the stellar luminosity is 4000 L$_{\odot}$. For a discussion on this in relation to the measured Hipparcos distances, see  \citet{Olofsson02}.

\subsection{SiO model}
The excitation analysis  includes radiative excitation through the first vibrationally excited state. Relevant molecular data are summarized in \citet{Delgado03b}.

The abundance distribution of the circumstellar SiO is of importance
in the radiative transfer modelling, and it is determined by
two processes, dust condensation and photodissociation, as the gas
expands away from the star. As outlined in more detail in 
\citet{Delgado03b}, the abundance is expected to decline, from the
stellar atmosphere equilibrium chemistry value,
once condensation of SiO onto dust becomes efficient when the kinetic
temperature drops below the condensation temperature. This requires a
minimum density which is not obtained in the low mass loss rate objects.
For the higher mass loss rates the abundance declines drastically
until the condensation ceases to be effective. Eventually, photodissociation
sets in and this limits the size of the SiO envelope. Thus, to a first
approximation the SiO abundance distribution consists of two components,
a compact high abundance region close to the star, and a more extended
low abundance region. The former dominates for low mass loss rates and
the latter for high mass loss rates. \cite{Delgado03b} used only one 
component and its size was estimated from a formula obtained by fitting
multi-line SiO data for eleven stars. 

In order to investigate which SiO abundance distribution that best reproduces the
flux picked up by the interferometer, the same ($u, v$) sampling was applied to the 
predicted brightness distribution from the model envelope. Two types of abundance 
profiles were used (by abundance we here mean the fractional abundance of
SiO with respect to H$_2$): i) a simple Gaussian abundance profile with an inner abundance 
$X_0$ and an $e$-folding distance $r_{\mathrm e}$ and ii) a Gaussian profile with a depleted abundance $X_{\mathrm D}$ and an $e$-folding distance $r_{\mathrm e}$, and an added high abundance inner component $X_0$ introduced as a step function inside a 
radius $r_{\mathrm c}$. This provides a first approximation to the scenario outlined above where a combination of adsorption of SiO onto dust grains and photodissociation dictate the abundance profile.

\begin{figure*}
\centerline{\includegraphics[width=17cm]{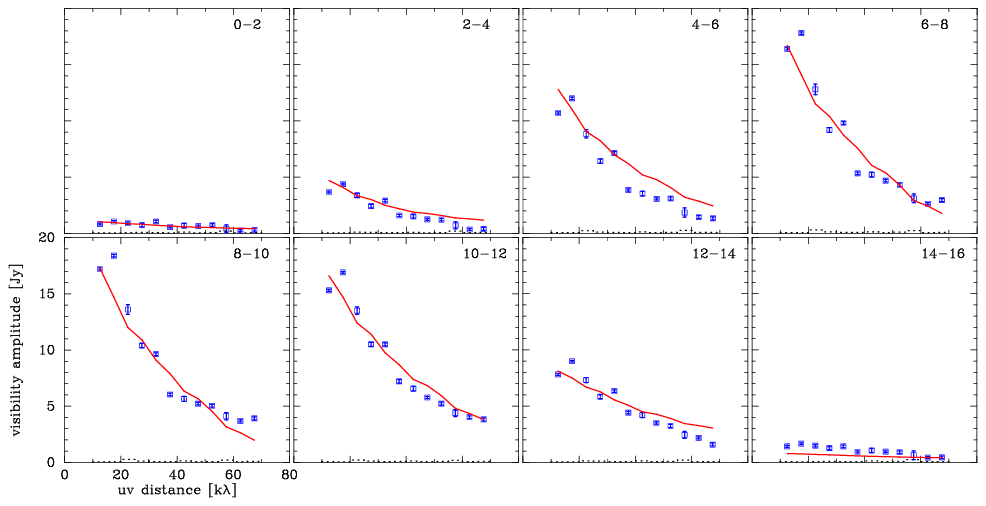}}
  \caption{Azimuthally averaged visibility amplitudes, in 5 k$\lambda$ bins,
across various parts of the line profile (as indicated in the upper right hand corner in km\,s$^{-1}$ given in the LSR frame) for \object{R Dor}.  The error bars indicate the 2$\sigma$ statistical error and does not contain the calibration uncertainty of about 15\%.
The solid line represents the best fit model using a constant expansion velocity.
The dotted histogram indicates the zero-expectation level, i.e., when no source emission is present.} 
  \label{rdoruvamp}
\end{figure*}

\subsection{R Dor}
\label{RDor:mod}
The SiO abundance distribution used by \citet{Delgado03b} ($X_0=5\times 10^{-6}$; $r_{\mathrm e}=3.3\times 10^{15}$ cm) 
fails to explain the observed visibilities on all baselines, see the dashed line in Fig.~\ref{rdor_velvis}. 
This indicates that the abundance is higher in the inner part of the envelope than what is given by the model. 
It should however be pointed out that the analysis of \citet{Delgado03b} is based on a size formula for the SiO
envelope which is obtained through multi-line fitting of data for eleven stars.
A better fit is obtained for $X_0=5.5\times 10^{-5}$ and 
$r_{\mathrm e}=1.0\times 10^{15}$ cm (dash-dotted line), but all the flux is not recovered on the shortest baselines using this model. Also, the fit to the single-dish data is worse. Instead, a combination of a high abundance compact component and a low abundance more extended component ($X_0=4\times 10^{-5}$; $r_{\mathrm c}=1.2\times 10^{15}$ cm;  $X_{\mathrm D}=3\times 10^{-6}$;  $r_{\mathrm e}=3.3\times 10^{15}$ cm) provides a very good fit to the interferometer data (solid line), and at the same time to the single-dish data as seen in Fig.~\ref{rdorspec}.

Of considerable interest is to investigate the assumption made in our models of a constant velocity throughout the SiO line emitting region. 
The observed line profiles are best fitted using a somewhat lower expansion velocity of 4.9 km\,s$^{-1}$ than obtained from the CO analysis (5.3 km\,s$^{-1}$). This suggests that the wind has already been accelerated to within $5-10$\% of its terminal value at radial distances of $\approx 1\times 10^{15}$ cm ($\approx 30$ stellar radii).
In Fig.~\ref{rdoruvamp} the model predictions are compared with the observations in eight velocity intervals. The model can account for the trend that the gradient in flux with baseline decreases towards the line wings. To test the sensitivity of the SiO line emission to the adopted velocity law a velocity gradient appropriate for a dust driven wind \citep[e.g.,][]{Habing94}
\begin{equation}
v(r) = \sqrt{v_{\mathrm i}^2+(v_{\infty}^2-v_{\mathrm i}^2)(1-\frac{r_{\mathrm i}}{r})}
\end{equation}
was first assumed. The ratio of the terminal wind velocity $v_{\infty}=5.3$ km\,s$^{-1}$ to the velocity $v_{\mathrm i}$ at the inner radius $r_{\mathrm i}=1\times10^{14}$ cm was fixed to 0.25. This velocity law provides a good fit to the line width of all available SiO and CO single-dish spectra. It was found that this velocity law only changes the results very marginally and provides an equally good fit to the observed visibilities as does a constant velocity field.  Models for pulsation driven winds \citep{Winters00b,Winters02,Winters03} generally suggest a more slowly increasing velocity with radius. A velocity law having a constant velocity of 5.3 km\,s$^{-1}$ outside $30$ stellar radii ($\approx 1.4\times 10^{15}$ cm) and 
\begin{equation}
v(r) =  5.3 \left( \frac{r}{1.4\times10^{15} {\mathrm{cm}}}\right)^{0.5} \ \ \mathrm{km\,s}^{-1}
\end{equation}
inside this region approximates such a slow increase. It provides equally good fits to the observations.
The reason that no significant difference is observed is related to the fact that
the SiO $v=0, J=2\rightarrow1$ emission is optically thick. Thus, the emission emanates from a small region close to $r_{\mathrm e}$ over which no large velocity gradient exist.

A point worth mentioning is that the relatively high turbulent velocity of 1.5 km\,s$^{-1}$ obtained for \object{R~Dor}, as required to fit the line shapes for both the CO and SiO line emission, also significantly affects the excitation of, in particular, the lower rotational levels. In the modelling it was found that the nominal value of 0.5 km\,s$^{-1}$ used by \citet{Delgado03b} results in population inversions in the lower level transitions of SiO, i.e., maser action. The line profiles in that case have strong peaks at the extreme velocities which is not observed. The higher turbulent velocity helps to quench the maser emission and produces line profiles that very closely resembles the observed ones, Fig.~\ref{rdorspec}.

\begin{figure}
\centerline{\includegraphics[width=8cm]{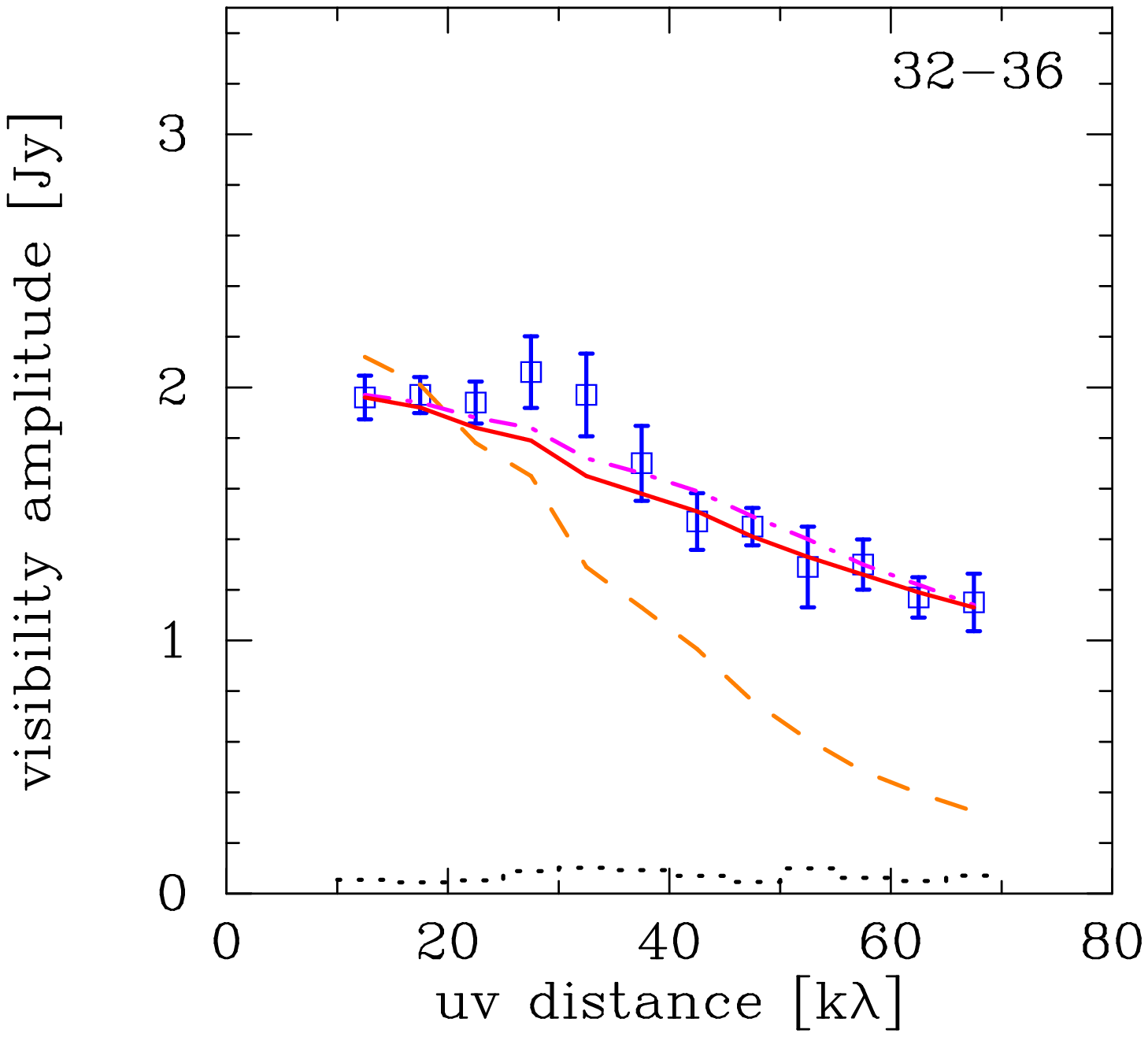}}
  \caption{Averaged visibility amplitudes, in 5 k$\lambda$ bins,
across the peak of the line profile ($32-36$~km\,s$^{-1}$ in the LSR frame) for \object{L$^2$ Pup}. 
The error bars indicate the 2$\sigma$ statistical error and does not contain the calibration uncertainty of about 15\%.
The solid line indicates the best fit model ($X_0=6\times 10^{-5}$; $r_{\mathrm c}=8\times 10^{14}$ cm;  $X_{\mathrm D}=2\times 10^{-6}$;  $r_{\mathrm e}=2.1\times 10^{15}$ cm), the dashed line the best fit model from \citet{Delgado03b} ($X_0=1.4\times 10^{-5}$; $r_{\mathrm e}=2.1\times 10^{15}$ cm), and the dashed-dotted line the best fit model without freeze-out ($X_0=5\times 10^{-5}$; $r_{\mathrm e}=1.1\times 10^{15}$ cm). The dotted histogram indicates the zero-expectation level, i.e., when no source emission is present.} 
  \label{l2pup_velvis}
\end{figure}
\begin{figure*}
\centerline{\includegraphics[width=17cm]{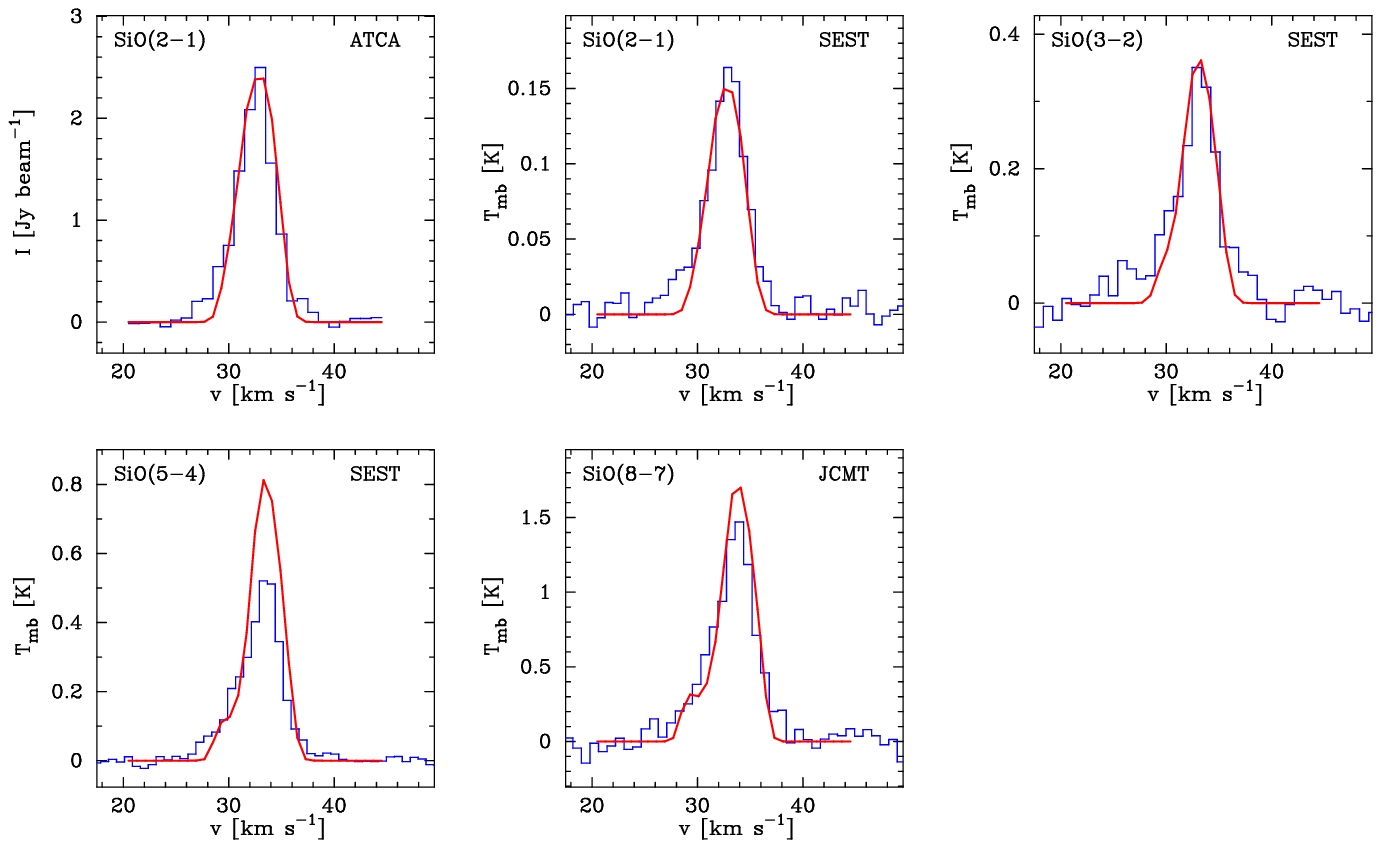}}
  \caption{Best fit model (solid line) overlayed on observations (histograms) in the form of single-dish SEST \citep{Delgado03b} and JCMT data (unpublished data) as well as the spectrum at the centre pixel in the cleaned ATCA map for \object{L$^2$~Pup}. The velocity scale is given in the LSR frame. The calibration uncertainty in the single-dish data is about $\pm 20$\%.} 
  \label{l2pupspec}
\end{figure*}
\begin{figure*}
\centerline{\includegraphics[width=17cm]{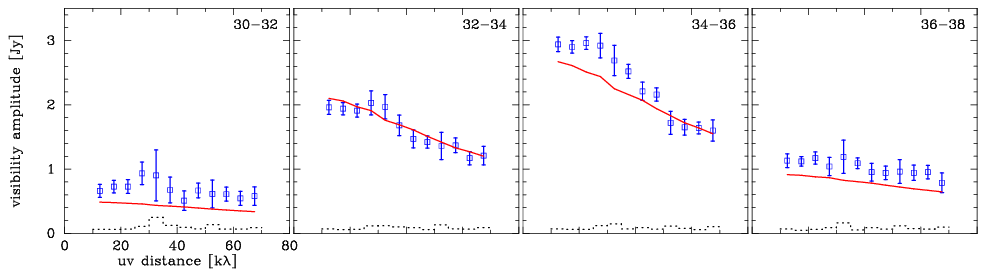}}
  \caption{Averaged visibility amplitudes, in 5 k$\lambda$ bins,
across various parts of the line profile (as indicated in the upper right hand corner in km\,s$^{-1}$ given in the LSR frame) for \object{L$^2$ Pup}. 
The error bars indicate the 2$\sigma$ statistical error and does not contain the calibration uncertainty of about 15\%. The solid line represents the best fit model using a constant expansion velocity. The dotted histogram indicates the zero-expectation level, i.e., when no source emission is present.} 
  \label{l2pupuvamp}
\end{figure*}

\subsection{L$^2$ Pup}

\object{L$^2$ Pup} has a significantly lower (a factor of $\approx 4$) mass loss rate than \object{R Dor}. However, the expansion velocity of the wind is also a factor of two lower resulting in only a factor of about two lower density in the wind. Thus, the SiO envelope characteristics should
be rather similar for the two sources.

The SiO abundance distribution used by \citet{Delgado03b} ($X_0=1.4\times 10^{-5}$; $r_{\mathrm e}=2.1\times 10^{15}$ cm) does not provide a good fit to the observed visibilities as illustrated by the dashed line in Fig.~\ref{l2pup_velvis}. While the model can account for the total flux, as indicated by the shortest baselines, it fails to account for all the flux at the longer baselines. This again indicates that the abundance is higher in the inner part of the envelope than what is given by the model. The flux seen by the interferometer at the longer baselines can be accounted for by raising $X_0$ to $4.5\times 10^{-5}$, at the same time that the size of the region needs to be lowered to $1.1\times 10^{15}$ cm in order to reproduce the total flux seen at the shorter baselines. This model is shown in Fig.~\ref{l2pup_velvis} as the dashed-dotted line and account very well for the flux at all baselines. It should be noted that this model provides a somewhat worse fit to the single-dish data, with a reduced $\chi^2_{\mathrm{red}}$ of 4.5. In particular the high-$J$ lines come out too strong in the model by about 50\% on average. In the light of the results obtained for \object{R Dor}, the combination of a high abundance compact component and a lower abundance more extended component ($X_0=6\times 10^{-5}$; $r_{\mathrm c}=8\times 10^{14}$ cm;  $X_{\mathrm D}=2\times 10^{-6}$;  $r_{\mathrm e}=2.1\times 10^{15}$ cm) produces an equally good fit to the interferometer data
(Fig.~\ref{l2pup_velvis}; solid line) and at the same time provides a much better fit to the available single-dish data ($\chi^2_{\mathrm{red}}=1.5$), Fig.~\ref{l2pupspec}. We note that our model can not reproduce the observed $J=5\rightarrow 4$ line at the SEST  which comes out with an integrated intensity of about 30\% higher than observed. The absolute calibration at the SEST is expected to be better than approximately $\pm20$\%.

In Fig.~\ref{l2pupuvamp} the model predictions are compared with the observations in four velocity intervals. The model can account well for the trend that the gradient in flux with baseline decreases towards the line wings. Note that the absolute scale of the flux in the model is somewhat lower in the wings. This is due to the fact that the model does not reproduce the line wings perfectly. Although, the source size estimated from the line wings is very well represented by the model. As for \object{R Dor} there is no evidence of any large gradients in the velocity field. Note that also for  \object{L$^2$ Pup}
the SiO $v=0, J=2\rightarrow1$ is optically thick and hence can not be expected to contain detailed information of the velocity law on smaller scales.

\section{Discussion}
\subsection{SiO abundance distribution}
\label{sec:abund}
Our modelling of the interferometer SiO data clearly show the need of an
abundance distribution with a high abundance ($X_0\approx 4\times 10^{-5}$), compact ($\approx 1\times10^{15}$ cm) component in order to reproduce the visibilities obtained with ATCA for both \object{L$^2$ Pup} and \object{R Dor}. In addition, there is evidence of a low abundance ($X_{\mathrm{D}}\approx 2-3\times 10^{-6}$),  more extended ($\approx 2-3\times10^{15}$ cm) component.

This scenario is very similar to the interpretation of interferometric SiS observations of the carbon star \object{IRC+10216} made by \citet{Bieging93}, where the abundance of SiS is seen to drastically drop by two orders of magnitudes at a radial distance of about $2\times 10^{15}$ cm.

The values of the initial SiO abundance derived here is very close to that expected from stellar atmosphere chemistry in  thermal equilibrium, $\approx 3-4\times 10^{-5}$  \citep{Willacy97,Duari99}. Assuming solar abundances a maximum SiO fractional abundance of $7\times 10^{-5}$ is obtained \citep{Delgado03b}.
 \citet{Duari99} further showed that non-equilibrium chemistry, where the chemistry is controlled by shocks, only has a very limited effect on the abundance of SiO in the inner part of the wind. Thus, it is expected from theory that the photospheric abundance of SiO remain relatively unaffected until the combined effect of adsorption and photodissociation sets in. Our data support such a scenario. This will be discussed further in Sect.~\ref{s:grainfor}.

As an alternative explanation to the low abundances of circumstellar SiO that
were estimated by \citet{Lambert78} and \citet{Morris77},
\citet{Scalo80} suggested that photodissociation by chromospheric radiation 
may be important. This is clearly not the case for the two sources studied here but a statistical sample needs to be investigated in order to rule out such a scenario. Unfortunately, not much is known about the chromospheres of AGB stars.

\subsection{Grain formation}
\label{s:grainfor}
The presence of a high initial SiO abundance followed by a drastic decrease indicates that adsorption onto dust grains could play an important role in removing SiO from the gas phase. \citet{Delgado03b} presented a relatively simple model to estimate how effectively SiO molecules freeze-out onto dust grains. 
Using this we find condensation radii of $\approx 5\times 10^{14}$ cm for both stars (essentially determined by
where the dynamical time scale equals the evaporation time). The degree of adsorption critically depends on the amount of dust available. For the dust parameters hinted at from the CO modelling and contained in the $h$-parameter we expect the SiO abundance to drop by about a factor of $\sim 3$ beyond this radius (for both stars). Likewise, we can use the simple
photodissociation model outlined in \citet{Delgado03b} to derive SiO photodissociation radii
of about $10^{15}$ and $3\times 10^{15}$ cm for \object{L$^2$~Pup} and \object{R~Dor}, respectively. Thus, within the considerable uncertainties, we expect
the SiO line emission to be strongly dominated by the compact high abundance region 
in \object{L$^2$~Pup}. In the case of \object{R~Dor} the low abundance region between
the condensation and photodissociation radii will also contribute. Our observational results
are certainly consistent with this, even though the estimated abundance drop beyond the
condensation radius is larger than predicted by the model. 
Given the simplicity of the treatment of the adsorption process and the uncertainty in
the adopted dust parameters, we find this acceptable.
More elaborate chemical models than the ones presented by \citet{Delgado03b} are needed to take the combined effect of adsorption and photodissociation properly into account. 

In addition, as indicated in Sect.~\ref{sec:abund} the role of chromospheric radiation as a means of dissociating SiO in these stars needs to be investigated. \citet{Scalo80} suggested that SiS might be the best probe of the degree of adsorption of Si bearing molecules since it can be readily formed though reactions with H and H$_2$ in the inner wind. In contrast there is no easy means of forming SiO. However, this was
based on the assumption that all Si-molecules were destroyed by chromospheric
radiation. Nevertheless, observations of SiS is an important complement to
the SiO observations, and it has been detected in O-rich envelopes \citep{Lindqvist88, Olofsson98b}.

\subsection{Wind dynamics}
The success of the modelling using a constant velocity (and a higher
micro-turbulent velocity width) strongly suggests that the wind has reached its terminal velocity already within $\sim  20-30$ stellar radii. This is further supported by the fact that the SiO and CO line data can be modelled using the same expansion velocity (within $5-10$\%). The marked difference in shapes between the SiO and the CO lines is attributed to strong self absorption of emission in the part of the wind moving towards the observer in the case of SiO.
Thus, we support the conclusion by \citet{Sahai93} rather than that of
\citet{Lucas92} where an extended region of acceleration was advocated, but
also note that we have so far studied only two low mass loss rate objects.

It has been suggested from hydrodynamical calculations \citep{Winters00b,Winters02,Winters03} that in very low mass loss rate  ($\dot{M}\lesssim 5\times 10^{-7}$ M$_\odot$\,yr$^{-1}$), low expansion velocity ($\lesssim 5$ km\,s$^{-1}$) objects, such as \object{R Dor} and \object{L$^2$ Pup}, the main driving mechanism behind these tenuous winds is stellar pulsation, and the wind acceleration is low. This is in contrast to winds which reach higher mass loss rates and expansion velocities where dust plays a primary role, and where the
wind acceleration is very efficient. From the present data we can not distinguish between these competing scenarios mainly due to the SiO emission being optically thick, and therefore most of the emission emanates from a region where the wind is close to the terminal velocity. Observations of optically thin SiS lines could possibly be useful to probe the velocity field. However, the SiS line emission is expected to be more than an order of magnitude weaker than that of SiO.

\begin{table}
\caption[]{Predicted SiO line intensities (integrated over the line and in main-beam brightness scale using a 12 m telescope) towards \object{R Dor} for selected rotational transitions.}
  \label{APEX}
$$
\begin{array}{p{0.15\linewidth}ccccccc}
\hline\hline
\noalign{\smallskip}
Transition & \nu & \theta_{\mathrm{mb}} & {\mathrm{Model\ 1}}^{\mathrm a} & {\mathrm{Model\ 2}}^{\mathrm b} & {\mathrm{Model\ 3}}^{\mathrm c}  \\
 & {\mathrm{[GHz]}} & [\arcsec] & {\mathrm{[K\,km\,s^{-1}]}} & {\mathrm{[K\,km\,s^{-1}]}} & {\mathrm{[K\,km\,s^{-1}]}}\\
\noalign{\smallskip}
\hline
\noalign{\smallskip}
$\phantom{0}5\rightarrow4\phantom{0} $       & \phantom{0}217.1   &                    29 & 20  & 19  & 24\\
$\phantom{0}8\rightarrow7\phantom{0}$        & \phantom{0}347.3   &                    18 &  28 & 33  & 39\\
$11\rightarrow10$                                                & \phantom{0}477.5   &                    13 & 30  & 46  & 55 \\
$16\rightarrow15$                                                & \phantom{0}694.1   & \phantom{0}9 & 28  & 58 & 63 \\
$20\rightarrow19$                                                & \phantom{0}867.5   & \phantom{0}7 & 27  & 65 & 63\\
$30\rightarrow29$                                                & 1300.5                      & \phantom{0}5 & 15  & 52  & 64 \\

\hline
\end{array}
$$
\noindent
$^{\mathrm a}$ Best fit model from \citet{Delgado03b} ($X_0=5\times 10^{-6}$; $r_{\mathrm e}=3.3\times 10^{15}$ cm).

\noindent
$^{\mathrm b}$ High abundance compact component ($X_0=5.5\times 10^{-5}$; $r_{\mathrm e}=1.0\times 10^{15}$ cm).

\noindent
$^{\mathrm c}$ Best fit model ($X_0=4\times 10^{-5}$; $r_{\mathrm c}=1.2\times 10^{15}$ cm;  $X_{\mathrm D}=3\times 10^{-6}$;  $r_{\mathrm e}=3.3\times 10^{15}$ cm).

\end{table}

\subsection{Predictions for high-$J$ SiO lines}
Presented here in Table~\ref{APEX} are estimated line intensities for selected transitions of SiO based on our models for \object{R~Dor} using a 12 m telescope. The differences between the one-component model derived by \citet{Delgado03b}, and the new models required to fit the interferometer data presented here, increase as higher  frequency transitions are considered. This shows the potential of 
single-dish telescopes such as CSO, JCMT and the upcoming APEX{\footnote{The Atacama Pathfinder EXperiment (APEX), is a collaboration between Max Planck Institut f\"ur Radioastronomie (in collaboration with Astronomisches Institut Ruhr Universit\"at Bochum), Onsala Space Observatory and the European Southern Observatory (ESO) to construct and operate a modified ALMA prototype antenna as a single dish on the high altitude site of Llano Chajnantor.}} and ASTE telescopes to study the properties of stellar winds close to the central star.
 
APEX and ASTE are of particular interest since they will eventually open up the THz region for observations. They will also be the only telescopes at the southern hemisphere being able to observe at sub-milimetre wavelengths.   

\section{Conclusions}
High resolution, interferometric, millimetre observations at $\approx 1\arcsec$ resolution of SiO $v=0, J=2\rightarrow 1$ line emission towards the two O-rich AGB stars \object{L$^2$ Pup} and \object{R Dor} have been performed. The emission is resolved, very centrally peaked, and suggests an overall spherical symmetry, even though there is an indication of a departure from 
spherical symmetry
in the case of \object{R Dor} at the arcsecond scale. A detailed excitation analysis was performed suggesting that the SiO abundance is very high ($\approx 4\times10^{-5}$) in the inner part ($\lesssim 1\times 10^{15}$ cm) of the circumstellar envelopes around both stars, consistent with predictions from LTE stellar atmosphere chemistry. For  \object{R Dor} the interferometer data further suggests that there is an additional, more extended, component with a significantly lower SiO abundance. 
We interpret this as the result of effective adsorption of SiO onto dust grains. 
Such a conclusion is less clear in the case of \object{L$^2$ Pup}.  

A comparison of model and observed line profiles further suggests that micro-turbulent motions are of the order of $1-1.5$ km\,s$^{-1}$. This is a significant fraction of the wind velocity in these slowly expanding winds ($\lesssim 5$ km\,s$^{-1}$), and it produces strong self absorption of the blue-shifted emission giving the SiO line profiles their characteristic shape. Additionally, the interferometer data provide constraints on the size of the region in which the wind is being accelerated. It is found that a constant velocity over the SiO emitting region can successfully account for the visibility amplitudes obtained at different velocity intervals. Furthermore, it is possible to model the line shapes of both SiO and CO emission using the same expansion velocity (within $5-10$\%) for both our sources. This constrains the acceleration region to within $\lesssim 20-30$ stellar radii. 

We conclude that SiO line emission plays a very important role in the study
of circumstellar envelopes,
both with respect to its dynamics and the gas and grain chemistry. However, it should be remembered that both sources studied here have very low mass loss rates and expansion velocities and that a more representative sample of O-rich sources needs to be studied in order to solidify the conclusion reached here. This could also be used to investigate the role of chromospheric radiation in photodissociation 
of SiO molecules. Such an effort is currently underway, and it justifies
the development of a more elaborate model to describe the abundance distribution of SiO.

\begin{acknowledgements}
FLS, HO and ML are grateful to The Swedish Research council for financial support.
TW is supported by an ARC-CSIRO Linkage Grant to the University of New
South Wales.
\end{acknowledgements}

\bibliographystyle{aa}

\end{document}